\documentclass[prd,onecolumn,a4paper,nofootinbib,11pt] {revtex4}
\usepackage{latexsym,amsmath,amssymb}
\begin{document}
\title{Some comments on ``The Mathematical Universe''}
\author{Gil Jannes \email{jannes@iaa.es}}
\affiliation{Instituto de Astrof\'isica de Andaluc\'ia, CSIC, Camino Bajo de Hu\'etor 50, 18008 Granada, Spain\\and \\Instituto de Estructura de la Materia, CSIC, Serrano 121, 28006 Madrid, Spain}
%
%
\begin{abstract}
I discuss some problems related to extreme mathematical realism, focusing on a recently proposed ``shut-up-and-calculate'' approach to physics~\cite{Tegmark:2007ud},\cite{Tegmark:2007bh}. I offer arguments for a moderate alternative, the essence of which lies in the acceptance that mathematics is (at least in part) a human construction, and discuss concrete consequences of this---at first sight purely philosophical---difference in point of view.
\end{abstract}
\maketitle
%
\section{Introduction}
\label{S:intro}
Modern-day fundamental physics has become increasingly mathematised, and this approach has been immensely successful in terms of describing and predicting some aspects of nature. Whether this implies an equally impressive advance in terms of {\it understanding} nature is a more delicate issue. According to the reductionist model, we understand certain phenomena of nature because we can describe them in terms of equations, which are justified in turn by `more fundamental' phenomena of nature, which can again be described by equations, etcetera. Alternatively, one could reject the reductionist chain based on the idea that, at every level of description and experience, complexity can emerge which cannot be reduced to the more `fundamental' layer of description.

The question is closely related to the debate between realist and constructivist views on science, i.e., the question whether the essence of science implies {\it uncovering} the elements of external reality and their relations, or rather {\it constructing} a framework whose truth-value is determined in part by human (for example, social) standards. A crucial argument in favour of scientific realism (the `no-miracle' argument) is that the success of science in describing nature would be miraculous if scientific theories were not at least approximately true descriptions of nature. Counterarguments include the underdetermination of theory by experiment, and the `pessimistic meta-induction' argument, namely that the historic fact that even immensely successful theories such as Newtonian gravity have been abandoned as candidates for `true descriptions' of nature leads one to expect that our current best theories will eventually suffer the same fate.

A view on science which combines a rather extreme form of scientific realism with an equally radical form of reductionism, is the idea to look for a most fundamental mathematical structure that `explains' it all, a `Theory of Everything' (ToE). Such a view could be dubbed `extreme mathematical realism', since the essence of the structure of external reality is assumed to be mathematical,\footnote{Mathematical realism in its plain form states that mathematical concepts exist independently of the human mind. When combined with the radically reductionist view that all elements of reality can be reduced to mathematical entities or structures, one arrives at what I here call `extreme mathematical realism', which is a form of what is known as `(radical) ontic structural realism' in the contemporary philosophical literature. I should stress that it is an issue of debate whether radical ontic structural realism necessarily reduces to a {\it mathematical} form of structural realism or not, and some authors hace criticised it precisely based on the argument that it does. See~\cite{ladyman-french} for ontic structural realism and how it could clarify identity and discernibility issues in quantum physics,~\cite{cao} for the critical argument that ontic structural realism implies mathematical reductionism and hence fails to account for physical entities, and~\cite{saunders} for a reply to this criticism.} and is arguably prevalent (at least implicitly) in a significant part of the high-energy physics community (see~\cite{weinberg} for an illustrative example,~\cite{zinkernagel} for some relevant case studies). A very interesting explicit formulation and defence of extreme mathematical realism was recently presented by Tegmark~\cite{Tegmark:2007ud},\cite{Tegmark:2007bh}. I wish here to discuss some aspects of extreme mathematical realism, and of Tegmark's proposal in particular, and point out what seem to me to be several problematic issues at the foundations of it. Far from swinging to the other, constructivist extreme (which, most scientists will agree, is clearly contradicted by the impressive success of science, and of mathematics' role in it~\cite{wigner}), I will advocate for an intermediate position. This consists essentially in the (mildly anti-Platonic) acceptance that mathematics is, at least in part, a human construction, without denying that its impressive success in science indicates that there is more to the relation between mathematics and reality than merely the fact that we use the former to describe the latter. I by no means claim this to be an original or very well-defined proposal, although I will try to stress points that are generally largely overlooked, for example the fact that whether or not mathematics is a human construction is (at least partially) an empirical question. Finally, I will point out that this---at first sight purely philosophical---argument has concrete consequences in the search for a possible Theory of Everything (although obviously `ToE' is a very badly chosen term once one agrees to dismiss extreme mathematical realism).

\section{Some problems of extreme mathematical realism}
\label{S:mathematical_realism}
Tegmark's proposal for extreme mathematical realism can be summarised as follows. He starts with the External Reality Hypothesis, or ERH, namely the hypothesis that ``there exists an external physical reality completely independent of us humans''. Second, he formulates the Mathematical Universe Hypothesis, or MUH, which states that ``our external physical reality is a mathematical structure,'' in other words that reality is really ``all just equations''. Tegmark argues that the MUH is actually a consequence of taking the ERH seriously, and therefore that to achieve a complete and objective description of reality, it must be formulated in a language completely independent of ``human baggage'': pure mathematics, without any reference to observed properties of physical objects. Moreover, according to Tegmark, this point of view makes (at least) two ``testable predictions'', namely that plenty of further mathematical regularities remain to be discovered in nature, and that parallel universes should exist in what Tegmark calls the ``multiverse level IV'' sense, namely that to every possible substructure of the mathematical superstructure which defines objective reality, there should correspond the physical existence of a universe.

\subsection{The External Reality Hypothesis}
\label{SS:ERH}
The question of the existence of an external reality has been a recurring theme in the history of philosophy. Two influential 20th-century examples are the following. In the philosophical current known as Phenomenology (see, e.g.,~\cite{husserl}, \cite{merleau-ponty} and \cite{heidegger}), it was held that whenever we talk about an `objective reality', we conceive it in a way essentially rooted in the perception that we, human conscious observers, have of it. For example, {\it blue} is not the wavelength of a light ray at a frequency of 450 nm, it is {\it our perception} of such light. The {\it frequency} of such a light ray is not an intrinsic property of it, but rather of the way we measure and describe it. Even its {\it being} a light ray is arguably not an intrinsic property of the light ray, but of our understanding of it. For example, a speculative non-human observer might categorise light rays as such based on their observational properties, not on any intrinsic non-observable property that essentially defines all light rays. In this phenomenological view, the claim that there exists an external reality {\it actually} independent of us humans is only meaningful inasmuch as the exclusion of human presence does not imply the exclusion of the {\it possibility} of conscious observation. On the other hand, Wittgenstein~\cite{wittgenstein2} defended the view that it is impossible to give conclusive rational arguments for the objective existence of an external reality, but that it is nevertheless an essential presupposition or background for any acquisition of knowledge, and hence for any serious philosophical or scientific debate.

The contemporary view can be seen as a combination of these two: While it is hard to define the precise meaning of (the existence of) external reality, generally speaking, its use is perfectly fine. Indeed, it is in general not necessary for there to be an actual observation to be able to speak of an observable, only potential observation is necessary. So, although the epistemological status of such an argument remains a bit questionable when literally expanded to reality as a whole (how should potential observation of an objective reality independent of conscious observers be interpreted?), it seems fine as a working hypothesis for science. Nevertheless, it might be wise to interpret a logical reasoning which is profoundly based on it, in particular: `{\bf {\it if} ERH {\it and} MUH, {\it then}  mathematical universe}', with care.

In any case, it is indeed clearly not too controversial that there is something out there which ``can be kicked, and can kick back''~\cite{popper}. But it must be emphasised that in this sense, the claim of the existence of an external reality relates to an observable physical reality (even when abstraction is made of any actual observation), to what one could call the Aristotelian reality of the senses, as opposed to the Platonic world of perfect triangles and other ideal (mathematical or metaphysical) concepts. In the opposite case, that is: if the ERH concerns the Platonic world, then the claim of its objective existence independent of us humans would certainly be far more problematic, and the whole argument for mathematical realism would be rather question-begging.

\subsection{The Mathematical Universe Hypothesis}
\label{SS:MUH}
The second element of mathematical realism is what Tegmark calls the MUH or `Mathematical Universe Hypothesis': the idea that our external physical reality {\it is} a
mathematical structure. I will sketch three arguments against this hypothesis. 

First, there is a possible confusion between what I have just called the Aristotelian observable world, and the Platonic world of (mathematical or metaphysical) ideas and concepts. The argument that the mathematical structure is the essential or fundamental one, and the implication that the observational content of the physical world is part of the baggage which should be thrown out to arrive at a Theory of Everything, seems to lead to a contradiction with the original ERH, since---as I argued above---the latter vindicated precisely the existence of an objective external {\it physical} and hence observable world, not of a mathematical one. Classical Platonism avoids this pitfall precisely by denying the objective existence of the observable world from the start, and calling it an illusion. Tegmark's way out of the contradiction is by arguing that the observational content ``emerges'' from the mathematical structure by a strictly mathematical analysis of the structure itself. This leads me to the second point of critique.

The second point is the question of how mathematical equations could arrive at meaning when human baggage is excluded. Tegmark argues that a purely mathematical investigation of the mathematical equation or structure can lead to the emergence of familiar physical notions and interpretation, and offers a possible first move in this direction,\footnote{An interesting second move to derive ``physics from scratch'' is suggested in~\cite{Bernal:2008qp}.} which places a heavy emphasis on the analysis of symmetries. The question is whether this could ever be sufficient to arrive at meaning, in the sense of ``human baggage'' or physically observable properties. An illustrative example is the following. In classical mechanics, all Hamiltonian systems with a single degree of freedom, with bounded trajectories and no explicit time-dependence, are integrable and mathematically equivalent, in the following sense. One can always define a canonical transformation such that these systems are described by exactly the same trajectories in the phase space of these new variables $Q$ and $P$. For example, using the Hamilton-Jacobi method, one can write $H=P$ and obtain $Q=t+Q_0$. This results in circular trajectories run through at a constant angular velocity, with the same conserved quantities of motion $P$ and $Q_0$ for {\it all} such systems. In particular, the difference between a harmonic and an anharmonic oscillator then depends solely on the choice of a `physical' coordinate system, since in terms of the canonical coordinates $Q$ and $P$, they are mathematically identical.\footnote{One could of course also define alternative sets of canonical coordinates in which the difference between a harmonic and an anharmonic oscillator is preserved, but this is rather trivial. The non-trivial fact is that a set of coordinates can be found in which an exhaustive mathematical solution (i.e., one accounting for all degrees of freedom) provides no way to discriminate between them. That this is not trivial is easy to see when one realizes that for two-(or higher-)dimensional oscillators, it is in general impossible to find such a coordinate transformation between a harmonic and an anharmonic oscillator, i.e., the distinction between the two is then intrinsic at the mathematical level and is independent of the actual choice of a physical coordinate system.} In other words, one and the same mathematical description results in two very different physical objects: a harmonic or an anharmonic oscillator, depending on the coordinate system that is chosen or imposed by the physical context. So the physical content of the object is in this simple example clearly not exhausted by its purely mathematical description.

The third point is the following. Tegmark argues that, if we want to give a complete description of reality, then we will need a language independent of us humans, understandable for non-human sentient entities, such as aliens and future supercomputers. However, it is not clear why we should recur to aliens or supercomputers. We know many non-human entities, plenty of which are quite intelligent, and many of which can apprehend, memorise, compare and even approximately add numerical quantities~\cite{dehaene}. Several animals have also passed the mirror test of self-consciousness. But a few surprising examples of mathematical abstraction notwithstanding (for example, chimpanzees can be trained to carry out symbolic addition with digits~\cite{dehaene}, or the report~\cite{parrot} of a parrot understanding a ``zero-like concept''), all examples of animal intelligence with respect to mathematics are limited to basic counting abilities. The point is that the question of whether mathematics is really an observer-independent language is in part an empirical question, not just a question of principle. This point is largely ignored by the contemporary philosophical discussion about the status of mathematical entities, which focuses mainly on epistemological and ontological arguments as to whether mathematical entities should be taken to really exist (as abstract, and therefore non-spatiotemporal objects, completely independent of us and our minds), as in the Platonic view, or to be, for example, useful but nevertheless `fictional' generalisations of certain properties of concrete physical entities.\footnote{There exists a whole array of possible positions with regard to the status of mathematical objects, see~\cite{balaguer} for a survey. The following example might be useful to illustrate the main contemporary arguments. `3' can be taken to either truly exist as an abstract entity, as in the Platonic view, or it can be taken to be merely a useful term for describing particular concrete sets of concrete physical entities, without implying the existence of the abstract entity '3'. In this latter view, `3' could then for example be considered as a fictional concept much in the same sense as, say, Sherlock Holmes. An important argument in favour of the existence of abstract objects (and hence in favour of Platonism), known as the `singular term argument', is essentially the following. It seems evident that the sentence '3 is prime' is true. However, in the traditional interpretation of truth, accepting sentences of the type `$a$ is $P$' as true implies a commitment to the existence of a referent for $a$. Russell's famous example `The present king of France is bald' should therefore be considered as false~\cite{russell}. This argument seems to imply that accepting the truth of `3 is prime' requires accepting the existence of the abstract entity `3'. The main argument against Platonism---apart from Occam's razor, see below---is currently held to be the  epistemological argument that it is questionable that we, as concrete spatiotemporal beings, could arrive at knowledge of mathematical objects if the latter were truly abstract non-spatiotemporal entities and therefore causally disconnected from us.} To settle the empirical question about the observer-independence of mathematical objects in the affirmative, non-human intelligent beings should exist that understand the language of advanced mathematics. However, none of the non-human intelligent beings that we know of confirm the status of (advanced) mathematics as an objective language. So maybe it is more reasonable to accept that mathematics is at least in part a human construction. Possibly, in Kronecker's words~\cite{biermann}, ``the natural numbers come from God, everything else is man's work''. From this point of view, the reality that we humans describe in mathematical terms is not an external reality independent of us humans, but rather an abstract entity which is defined (at least in part) by us, humans.

\section{A moderate alternative and its consequence}
\label{S:moderate}
The alternative to the extreme mathematical realism defended by Tegmark simply consists, as I just indicated, in the acceptance that mathematics is at least in part a human construct. An interesting side-question is whether such a general point of view can have any concrete consequence. The answer is affirmative, at least in the sense of denying (part of) the concrete ``testable consequences'' of the extreme mathematical realism defended by Tegmark. 

First of all, I should emphasise that mathematics is `at least in part' a human construct. Indeed, it makes equally little sense to claim that mathematics is {\it nothing more} than a human construct as it is to claim that mathematics is the true and pure language of reality. Therefore, my contention about mathematics and its (partially) human origin should not be understood in the sense of intuitionism or constructivism (which claim that mathematical entities are essentially mental constructions), but simply as a rejection of Platonism, at least in its traditional formulation.\footnote{In~\cite{tait}, the thesis is defended that the question of whether mathematical entities really exist is not only irrelevant for mathematical practise, but even for Platonism, when this is understood in a deflationary or moderate sense.} A classical argument against Platonism is that it unnecessarily complicates our view of reality by requiring a commitment to the existence of an immense realm of mathematical and other abstract entities, thereby violating Occam's razor. This does not imply a denial of the effectiveness of mathematics at describing (part of) reality. But the reason for this effectiveness might well lie simply in the fact that mathematics is a very flexible tool, an abstract descriptive language which---precisely because it is {\it abstract}---can be applied to a huge variety of different {\it concrete} situations.

Let me now discuss consequences of this view. Tegmark predicts the existence of plenty of further mathematical regularities to be discovered in nature, and the existence of a ``level IV'' multiverse. 
The first prediction would also be the case if mathematics is interpreted simply as an abstract descriptive language, provided that it is sufficiently flexible to describe a wide variety of concrete situations. Since `old' mathematics (differential equations, for example) are constantly being applied and used to describe `new' situations or applications (in economic or demographic models, for example, where it seems hardly contentious that the use of mathematics is descriptive and not fundamental), this is clearly the case. Second, in the line of the first argument against the MUH (see section~\ref{SS:MUH}) on the confusion between mathematical and physical existence, the fact that many mathematical structures might exist, either totally separate or as substructures of one unique mathematical superstructure, does {\it not} necessarily imply the physical existence of a multiverse. 

Apart from rejecting these two predictions of extreme mathematical realism, there is a third consequence of accepting that mathematics is a descriptive and partly human language. In extreme mathematical realism, the ultimate unifying physical theory must necessarily be the unique mathematical superstructure, since reality {\it is} precisely this mathematical superstructure. If one rejects mathematical realism, then there are two possibilities. The pessimistic possibility is that mathematics is not flexible enough, or our use of it not imaginative enough, to describe scientific reality at the fundamental level of time, space and matter in a single comprehensive framework. Then we simply will not be able to define a single correct ToE. The optimistic option is that mathematics should be capable of describing any sort of (existing or non-existing) universe or multiverse, for many of which it might be hard to imagine any meaning in human or physical terms, in spite of their mathematical consistency. The selection of which theory is {\it the} correct ToE should then be based not only on a criterion of mathematical consistency, but at least also on additional experimental input. And even that might not be sufficient. At the level of quantum gravity, the lack of any direct experimental input might make underdetermination of theory by experiment~\cite{quine} an acute problem, since the further an experiment is from our everyday reality (for example, due to the extremely high energies involved), the further the involved concepts are from directly identifiable physical observables. There might then be several, maybe even countless, abstract mathematical structures which, in the proper limit and with the proper interpretation, give rise to time, space and matter as we know it. In that case, in the end maybe criteria of elegance and other human preferences might come into play to determine the `correct' ToE.

\section{Conclusions}
\label{S:conclusions}
I have discussed a few arguments against extreme mathematical realism, focusing on the formulation recently given by Tegmark~\cite{Tegmark:2007ud},\cite{Tegmark:2007bh}. First, I pointed out a possible logical problem in identifying physical and mathematical existence. Second, I gave a simple example showing that physically observable properties cannot always be deduced from mathematical properties. Third, I emphasised that the available empirical evidence reinforces the anti-Platonic position that mathematics is, at least in part, a human construction. While none of these arguments is conclusive in itself, they illustrate that extreme mathematical realism implies a radical and therefore tense position (and possibly even an internal contradiction) in several long-standing philosophical debates. Finally, I have argued that rejecting extreme mathematical realism has concrete consequences in the search for a Theory of Everything, namely that there is no reason to expect mathematical consistency, or the search for a mathematical superstructure, to lead to a single---and therefore necessarily correct---Theory of Everything.

These arguments can be related to the general problems of ``radical ontic structural realism,'' the contemporary name for the philosophical current to which extreme mathematical realism belongs. Radical ontic structural realism is basically the idea that only structure in the sense of {\it relations} is real, and denies the reality of the objects or {\it relata} that instantiate these relations in the physical world. The basic problem~\cite{esfeld-lam} with such a radical view is that concrete, physical relations cannot exist without relata. A moderate form of structural realism would therefore accept that physical relations require relata, i.e., that the realisation of abstract mathematical structures requires real physical objects to instantiate them, while it could still defend that the intrinsic properties of these objects are exhaustively described by the structural relations between the objects. However, it is essential to understand that this does not reduce the concrete existence of such objects to the abstract existence of the relations: it does not imply that all possible (mathematical) relations are actually (physically) instantiated, a point related to my first argument. Thinking of the relata as the concrete objects that we encounter from experience or observation, and their relations as structures described by human-independent abstract mathematics, is a serious oversimplification, as illustrated by my third argument. Nevertheless, the essential point is that science cannot do without either of the two, and that neither of them can fully be reduced to the other, as I illustrated in my second argument. From this point of view, identifying physical and mathematical existence is an ontological category mistake, or at least a contentious metaphysical position. The same is therefore true for the prediction of a multiverse. To interpret this prediction as a prediction about physical existence requires not only a (mathematical) theory which predicts the (mathematical) existence of multiple structures, but also metaphysical premises of radical realism and reductionism. The other prediction of extreme mathematical realism, that plenty of further mathematical regularity remains to be discovered in nature, can just as easily be accomodated in the view that mathematics is an abstract descriptive language. 

This brings us back to an issue that I have mentioned in the introduction, namely the argument between reductionism and emergence. Extreme mathematical realism comprises a defence of reductionism driven to its most radical consequences.\footnote{For the sake of clarity, I should insist that this is not necessarily the case for (moderate versions of) scientific realism and reductionism in general. Scientific realism does not necessarily entail reductionism, nor does reductionism necessarily imply scientific realism.} As Tegmark correctly insists upon, symmetries have played a paradigmatic role in contemporary physics. However, this is true both on the reductionist front and in the study of emergent phenomena. Moreover, it is not only the symmetries themselves that determine our physical experience of reality, but also the way in which they are broken (as already insisted upon in Anderson's famous paper~\cite{anderson}) and/or formed. It is for example widely expected that the Brout-Englert-Higgs mechanism of `spontaneous' symmetry breaking is responsible for mass acquisition by the particles of the standard model when the energy decreased below a certain threshold. On the other hand, from the study of low temperature physics, we know that not only symmetry-breaking occurs when decreasing the energy, but that new symmetries can also emerge in that same process of energy decrease. Actually, there are several examples, for instance of condensed matter systems, where Lorentz symmetry emerges as an effective low-energy symmetry~\cite{Barcelo:2005fc}, and there are speculations (although no experimental evidence as yet) that this might also be the case for our spacetime as a whole~\cite{Volovik:2003fe}. It might then well be that the symmetries which we observe are a contingent result of the energy window in which we live, and all the physics that we deal with might be limited by the upper and lower cut-off scales at which the `fundamental' symmetries that we experience emerge and are broken. It would then be impossible to derive the full mathematical bird's point of view for our universe from our frog's point of view. In any case, since we have currently only approximately understood less than 5\% of the total energy content of our observable universe, there remains a lot of work to be done both on the reductionist and on the emergent front. But in the meantime, it might be well to keep an eye open on both. 
To end with Einstein's words~\cite{einstein}: ``As far as the laws of mathematics refer to reality, they are not certain; and as far as they are certain, they do not refer to reality.''


\section*{Acknowledgements}
I would like to thank Carlos Barcel\'{o}, Geert Engels, Luis Garay, Paul Jannes, Dani Paredes and Joachim Willems for valuable comments and discussions.


\end{document}